# Magnetism induced by vacuum polarization at the gold-monolayer interface

Itai Carmeli

Abstract

Giant magnetization of polyalanine monolayers on gold observed in resent years along with unusual large spin selectivity, the potential of the film and its temperature dependence are all explained in the context of vacuum polarization at the gold-monolayer interface. The enhancement is directly related to the fine structure constant, $\alpha$ and involves a change in basic physical parameters including changes in the permeability of the vacuum and in the local zero point energy (ZPE) of the gold-monolayer interface. Formation of an ordered state by the self-assembly process and the interaction of the monolayer with the gold plasmons form a new state of matter that can interact strongly with the ZPE.

New emerging experimental results[1-7] show that when diamagnetic organic molecules are adsorbed on diamagnetic gold surfaces new magnetic properties arise. The magnetic moment of the system is very large and can reach values of up to hundred Bohr magnetons per adsorbed molecule[1], it is temperature independent and demonstrates large anisotropy with the magnetization pointing along the molecule long axis, perpendicular to the surface. The magnetic properties depend on the level of organization and quality of the film. Among this observation there are increasing numbers of reports which indicate that this new type of magnetism is not specific just for interaction of gold surfaces with organic molecules but can be found in other types of systems. Similar type of magnetism has been observed in the dielectric material HfO2[8], in etched silicon[9], DNA[4,7], Lewis conjugates[5] monolayers on gold surfaces and in thiol-capped gold nanoparticles[6]. There is a common denominator underlying this phenomenon; in all these systems magnetism occurs at the surface of the material, it is anisotropic and is temperature independent. The question than asked, what is so unique about surfaces that render them to this new type of magnetic properties. The purpose of this manuscript is to suggest general guidelines and to draw attention to an alternative way of looking at this new and unique type of magnetism. A more rigorous formulation is outside the scope of this paper and will be the subject of further studies.

It is quite well known that the properties of surfaces vary from the bulk. Recently it has been demonstrated that when diamagnetic molecules are adsorbed onto gold surfaces new electronic properties of the surface emerge which are due to electron transfer between the adsorbed monolayer and the surface, resulting in charge rearrangement at the surface level. This charge rearrangement is observed as a new band in the absorption spectra of the gold/monolayer surface and is attributed to delocalized electrons, suggesting the appearance of plasmon or exciton like behavior of the surface electrons[10,11]. It will be argued that this new band is responsible for the magnetic behavior observed for molecular monolayers on gold surfaces. Very interestingly, it was recently observed that thin (nm range) metallic films[12], nanometer sized silver rings[13] and metallic nanostructers[14] exhibit diamagnetism which is about 1-3 orders of magnitude larger than that of the bulk metal. This indicates to some sort of mechanism which enhances the inherent diamagnetism of the metal and occurs at the surface. The new type of magnetization is diamagnetic in nature and temperature independent. It originates from zero point energy (ZPE) contributions to the plasmon field and can be seen as additional vacuum contribution to the magnetic moment of the metal sample. The plasmon field gives rise to a large change in the surface ZPE and vacuum pressure[14-16] due to interaction of the vacuum modes with the plasmons in the metal. This interaction is predicted to produce large electric polarization, changes in the electric field energy[16] and in the magnetic properties of the surface[14] as has been confirmed experimentally[12-14]. Unlike the positive magnetism observed for organic layers on gold, this type of magnetism is diamagnetic. However it will be argued that both types of magnetism emerge from plasmons which enhance the magnetic response of the material. This paper will attempt to show that the two observation of the change in ZPE, the emergence of giant positive magnetization of thin organic films on metal surfaces and the large increase in demagnetization of thin metallic films are tightly connected.

Experimental results addressing the spin selectivity of polyalanine monolayers on gold surfaces, the magnetic moment of the films and the temperature dependence of the surface potential will be discussed. The experimental results will be shown to diverge from the predicted values obtained by simple theoretical considerations. These considerations assume that the electronic properties of the molecules composing the monolayer are identical to the electronic properties of the isolated molecules prior to adsorption (assumption A). It is proposed here that the discrepancy between the experimental observations and theory (based on assumption A) is due to interaction of the system with the vacuum which produces a change in the local ZPE. Due to this interaction the electronic properties of the gold/monolayer interface change. The electronic structure can no longer be considered as an assembly of the individual electronic properties of the isolated molecules but as a new phenomena resulting from collective interactions between the molecules of the monolayer, the gold surface and the vacuum.

In order to obtain the new physical properties of the gold-monolayer interface that give rise to the observed enhancement in magnetism a new constant will be introduced. This constant is proportional to the inverse of the fine structure constant- $\alpha$. By using a scaling function for electrons orbiting around a nucleus the new electronic properties of the layer are obtained. It will be shown that by applying the scaling procedure to the theoretical consideration based on assumption A, an exact match with the numerous experimental observations is obtained.

We assume that the interaction of the gold/monolayer interface with the vacuum causes a change in the local ZPE. The change in ZPE will redefine in terms the basic electronic properties of the system. Due to interaction of the interface electrons with the vacuum there is a reduction in their ZPE which causes an increase in the electrons de Broglie wavelength and to their expansion. Using a scaling function for electrons orbiting around a positive nucleus[17] we can define the new physical parameters caused by the interaction of the gold-monolayer interface with the vacuum. The parameter $r_k$ (eq.1) when multiplied by a constant (q) satisfies the following equation of motion for electrons orbiting around a positive nucleus:

$$r_k'(t) = q^2 r_k(t/q^3) \text{ (eq.1)}$$

Where $r_k^{'}(t)$ is the new trajectory for the orbiting electron which is deduced by multiplying the trajectory of the electron former to the formation of the monolayer by the constant $q = \sqrt{\alpha^{-1}/2} \approx 8.3$ (eq.2), where $\alpha$ is the fine structure constant. The value of q defines the new radius (or the de Broglie wavelength) of the electrons caused by the interaction with the ZPE and thus will redefine the energy, momentum and all other physical properties of the gold-monolayer interface (eq 3-12). The parameter q can accept any value but the specific value defined for the parameter q was chosen in order to give an exact match between theory and the numerous experimental observations. The physical properties of the gold/monolayer interface will be obtained by replacing the units in the equations based on assumption A, with the following units obtained by the scaling procedure:

(3) $x' = q^2 \cdot x$   new distance

(4) $v' = q^{-1} \cdot v$   new velocity

(5) $c' = q^{-1} \cdot c$   new speed of light

(6) $F' = q^{-4} \cdot F$   new force

(7) $a' = q^{-4} \cdot a$   new acceleration

(8) $t' = q^3 \cdot t$   new time

(9) $h' = q \cdot h$   new plunk constant

(10) $E' = q^{-2} \cdot E$   new energy

(11) $V' = q^{-2} \cdot V$   new potential

(12) $E_{el}' = q^{-4} \cdot E_{el}$   new electric field

The unsubscribed letters represent the units used in the equations based on assumption A. The subscribed letters represent the new physical parameters of the gold-monolayer interface caused by the interaction with the vacuum. Presented bellow is a summery of experimental results and the way in which they diverge from the expected theoretical values based on assumption A:

1. The selectivity of the film to spin polarized electrons is very high and can reach values of up to 70% spin selectivity[2]. The asymmetry (A) is three orders of magnitude higher than that obtained and calculated in the gas phase. Using the fine structure constant we obtain the selectivity of the film:

$$A_{film} = A_{gas} \cdot (\alpha^{-1}/2)^2 \approx 1 \quad (eq.13)$$

2. Considering the large dipole moment the isolated molecule possess of about 55 Debye the potential at the surface of the monolayer should have been[18] $V_{sur} = |4.7|$ V. Applying the parameter q according with eq.11 gives a reduction of the surface potential by a factor of 68.5:

$V_{sur} = |4.7| V/q^2 = |0.07| V$ (eq. 14) which explains the low surface potential observed experimentally, a more exact treatment will be given later.

3. The dependence of the surface potential on temperature (will be discussed later).

4. Assuming about one unpaired electron per molecule as verified experimentally[10], magnetometer measurements revile susceptibility and magnetic moments which are a factor $q^3$ higher than values expected from a single unpaired spin per molecule.

The electronic structure of the gold/monolayer interface will be described as a two dimensional electron gas in spite of the fact that previous to formation of the monolayer the electrons in the isolated molecules are confined to the nucleus with an orbital radius of: 0.48Å (radius of an electron orbiting an oxygen atom). The canonical partition function in this case is:

$$Q = \frac{S^N}{4\pi h^{2N}} \cdot \frac{1}{N!} \cdot \left(\frac{4 \cdot 2\pi m_e}{\beta}\right)^{1N} = \frac{S}{4\pi (\Lambda/2)^2} \quad (eq.15)$$

Where: N is the number of particles, for simplicity, will be taken as one (one electron per molecule), $m_e$ is the electron's mass, S the surface area of the system were the electron is confined within and $\beta=1/KT$.

The de Broglie wavelength at 297K is:

$$\Lambda = \sqrt{\frac{h^2}{2\pi m_e kT}} = 4.3 \, nm \quad (eq.16) \text{ and the energy is:}$$

$$\langle E \rangle = -\left(\frac{\partial \ln Q}{\partial \beta}\right) = kT = 26 \, meV \quad (eq.17)$$

We want to compare the partition function of an electron bound to move at a radius of 0.48Å attracted by the Coulomb force of the nucleus in an isolated molecule ($Q_{0.48}$), to a free electron in the monolayer ($Q_{21.5}$) which has escaped the Coulomb attraction of the

nucleus and now possess a wavelength of 43Å. Assuming surface radius (S) of 0.48Å in both cases, we obtain:

$Q_{0.48Å}/Q_{21.5Å} = (4\pi r_{0.48}^2/4\pi r_{0.48}^2)/(4\pi r_{0.48}^2/4\pi r_{21.5}^2) = 1/5 \cdot 10^{-4} = 2 \cdot 10^3$ (eq.18).

Thus, the probability of finding an electron away from the nucleus is three orders of magnitude higher in the monolayer than at the isolated molecule, a result of the interaction of the gold monolayer interface with the vacuum which leads to a decrease in energy (eq.10) and thus the increase in the wavelength (eq.3) of the electrons. This implies a decrease in the energy uncertainty of the monolayer electrons and an increase in coherency throughout the system. A system where the electrons have a higher de Broglie wavelength in spite the Coulomb attraction of the nucleus indicate that the electron momentum has reduced, angular momentum increased, and the energy of the ground state has lowered. The electronic wave function is now more coherent, the electrons energy is closer too room temperature energy and can be coupled to the electronic structure of the gold surface.

The molecule polyalanine possess a large dipole moment which results from the separation of half an electron charge between the C-terminal (negative pole) and N-terminal (positive pole) of the molecule, a distance of 30Å. When polyalanine molecules were adsorbed onto a gold surface with the C-terminal binding to the gold, a well organized layer was formed[1-3,10]. Upon adsorption a dipolar layer was created with a residual negative pole pointing away from the surface, as indicated by contact potential difference (CPD) measurements[2]. This is curious since the intrinsic dipole moment of the molecule points to the opposite direction. Due to the large dipole moment of polyalanine molecules it is clear that in order to form a well organized layer intensive charge rearrangement must take place within the molecules of the layer and the gold interface, otherwise the large dipole dipole interaction would not allow the process of self assembly to take place[2]. The intense charge rearrangement is responsible for the reduction of the molecules dipole moment and the net residual negative pole. The potential across the layer ($V_{CPD}$) based on assumption A and according with reference-18 would have been -4.7V, for a layer thickness of 30Å, dielectric constant ($\varepsilon_r$) of 2.5, molecular surface area of $3.8 \cdot 10^{-19}$ m$^2$/molecule, a dipole moment of 55 Debye and taking into account pin holes in the layer which reduces the surface potential by a factor of 3 (discussed later). Experimental measurements of polyalanine monolayers on gold[2] indicated a $V_{cpd}$ of +0.28V instead of the calculated -4.7V. This difference is attributed to charge rearrangement that causes a reduction in the layer dipole moment and a reversal of its sign.

We consider a scenario in which the reduction of the dipole moment results from a transfer of half an electron from the gold to the monolayer. The electron will be transferred to the oxygen atom which is the most electronegative atom in the molecule and is located in close proximity to the surface. This electron transfer will initiate the formation of a new plasmon band that interacts strongly with the vacuum, which will further cause the expansion of the electron cloud due to a reduction in the ZPE. The expansion is from the C-terminal towered the N-terminal and the expanded electron leaves behind a hole in the electron distribution of the gold electrons. The degree of expansion will be determined by the scaling factor- q. The new distance of the electron 32.5Å is calculated by taking the orbital radius of polyalanine oxygen electron 0.474Å (which is close to the theoretical value 0.48Å) and multiplying it by the scaling factor $q^2$ (new distance using eq.3). At this point the diameter of the electron cloud or its wave length will be 64.9Å. It will cancel the original dipole moment of the molecule and will add to it an opposite dipole moment of +0.28V. The new value, which match the experimentally CPD obtained indicate that the positive pole of the layer dipole moment is now pointing towards the surface.

The driving force for the expansion is the strong dipole-dipole interactions between the molecules. Yet, there must be another state with low potential energy that will accept the electron in its new expanded state. This state arises from the total reduction in the ZPE of the gold-monolayer interface due to the new plasmon band that is formed. It is as if the electron is tunneling through two barriers. The first barrier is the Coulomb potential of the nucleus and the second is the thermal potential caused by thermal fluctuation of the surrounding. Tunneling through the first barrier the electron detaches from the Coulomb potential of the nucleus and achieves its free electron gas state with a diameter of 61Å, the wavelength of ½ an electron in a two dimensional free electron gas at 297K. Tunneling through the second barrier it reaches its final diameter of 64.9Å. The mechanism that allows the detachment of the electron from the nucleus is the reduction in the electronic energetic state by a factor $q^2$ as predicted from equation-10. The CPD data indicated that a dramatic change has occurred in the electrons wave function otherwise the value obtained for the CPD would be high and negative (-4.7V). It is proposed that this change is related to electron expansion. As the electron expands the positive charge of the hole left in the gold electron distribution increases accordingly. When the wave length of the ½ expanded electron reaches $\lambda_D$= 61Å the charge of the hole equals a positive value of +½ electron charge and a wavelength of $\lambda_D$= 61Å, the wavelength of ½ a hole in a two dimensional free electron gas at 297K. During the expansion the wave function of the ½ electron mixes with the wave function of the hole to form an exciton which consists of +½ positive charge and a -½ expanded molecular electron charge. Due to the mixing of the two electronic states the two systems (the gold and the monolayer) are integrated. The accumulation of positive charge as holes in the d-orbital of the gold electrons conduction band has been recently supported experimentally[6]. At $\lambda_D$= 61Å the wave lengths, the energetic state and the charge of the two particles match, the total charge of the exciton is zero and the potential of the layer will be zero ($V_{CPD}$=0). But due to the scaling factor the expansion propagates and the energy of the expanded electron is lowered bellow the Fermi energy of the gold electrons. To equalize the Fermi energy of the two systems (gold and the monolayer) there is charge transfer from the gold to the monolayer accumulating additional negative charge on the layer and increasing the positive charge of the hole. Due to this additional charge transfer the CPD obtains its positive value[2]. Taking the ratio of the wavelengths between the positive charged hole of the gold and the free negative molecular expanded electron we obtain: 61Å/64.9Å=0.94. Keeping in mind that this difference in wavelengths signifies the amount of charge transfer from the gold to the monolayer we can calculate the potential of the layer. The original half electron charge at the C-terminal of the molecule was expected to give a CPD reading of -4.7V. Taking (1-0.94)·100 = 6% of the original CPD value and reversing its sign gives +0.28V, a value which matches the experimental positive CPD value obtained.

To show that this concept works in a more quantitative way, we can use the following equation which calculates the potential of the film, taking into account the thickness and charge density on the monolayer:

$$V_{CPD} = D\sigma e/\varepsilon\varepsilon_0 \quad (eq.19)$$

Where D is the thickness of the film, $\sigma$ is surface free charge density (charge per unit area A), e is the elementary charge, $\varepsilon$ is the dielectric constant of the film and $\varepsilon_0$ is the permittivity of the vacuum. The layer thickness is 30Å with a dielectric constant of 2.5. There are $n_e=1.3 \times 10^{13}$ excitons/sample (discussed in later section) for a sample surface area of 0.15(cm$^2$). Thus $\sigma = n_e/A = 1.3 \times 10^{13}$(excitons/sample)/$1.5 \times 10^{-5}$(m$^2$) $= 8.7 \times 10^{17}$ excitons/m$^2$. To calculate the negative charge of the layer due to the charge transfer we multiply the number of excitons by 0.06, the ratio between the wavelengths of the positive and negative charges of the excitons, and divide by 2 since the ratio is between ½ charged particles. We obtain a total negative charge density of $2.6 \times 10^{16}$ electrons/m$^2$. Substituting these numbers in eq.19 we obtain a positive potential of $V_{CPD}= +0.56V$. Since the charge is uniformly distributed over the thickness of the layer the center of mass of the film electronic distribution is half the layer thickness i.e. 15Å. Substituting this number in eq.19 we obtain the experimental CPD value and the previously calculated potential of +0.28V. The positive sign of the CPD is due to the film being negatively charged. Equation 19 is used to calculate the potential drop across a plate capacitor. An alternative way to look at the gold-monolayer interface is as a capacitor. The gold side is positively charged with a charge density of $2.6 \times 10^{16}$ holes/m$^2$, the layer is negatively charged with a uniform charge density of $2.6 \times 10^{16}$ electrons/m$^2$. Since the negative charge is not all concentrated 30Å away from the gold surface but uniformly distributed across the monolayer it is justified to take 15Å as the center of mass for the electronic charge distribution.

The next step is to explain the CPD dependence on temperature. Experimentally we observe that around 264K the CPD nulls and at 250K becomes negative[2]. As the temperature decreases the de Broglie wavelength of the gold hole expands. Using equation No.16 we calculate the wavelength of the hole at 264K to be 64.9Å. That accounts to a 3.9Å increase in wavelength on the decrease of temperature from 297K to 264K. Thus, at 264K the wavelengths of the positive and negative particles of the pair overlap, creating a resonance between the gold and the monolayer electrons and nulling the potential of the surface. The reduction in the surface potential is due to charge transfer from the layer back to the gold because of the decrease in Fermi energy of the gold electrons. Again, the ratio between the wave lengths of the two particles determines the charge on the layer. Not only that the wave lengths of the two particles coincide at this temperature, in addition, the thermal energy $E_{KT}$ that couples the particle to it's surrounding, in our case to the gold electrons, matches the electrostatic energy ($E_{el}$) which is the Coulomb energy between the proton and the ½ expanded electron. The matching in the wave length and energies leads to intense resonances as observed in the electron transmission spectrum[2]. Indeed, calculation show that the kinetic energy $E_{KT}$=22.7meV almost equals the electrostatic energy $E_{el}$=22.2meV at 264K. The calculation for the electrostatic energy was performed using the Coulomb equation with a dielectric constant of 2.5. Taking into account the experimental error the two values actually overlap.

The resonances obey a particle in a box equation:

$$E_n = \frac{n^2 h^2}{8mL^2} \quad (eq.\ 20)$$

Fitting to the energy of the most prominent pick[2] at $E_1$= 18±2mV (n=1) and substituting for m= ½$m_e$ we obtain that the size of the box L=65±4Å match surprisingly well the wavelength at which the particle pair coincides and the intense resonance occur. As the temperature is further lowered the resonances disappear[2] and the CPD changes in sign. Applying the same logic we calculate the CPD value at 250K to be -0.12V, which match the experimental value obtained. Thus, this model which assumes the creation of an exciton pair at the gold-monolayer interface due to the reduction in the ZPE of the gold-monolayer interface gives a good prediction for the CPD value and sign, its dependence on temperature (switching signs) and the temperature where its value nulls and intense resonances appear in the electron transmission spectra.

The physical picture emerging so far is of a two dimensional free delocalized exciton or plasmon gas formed by the mixing of the expanded ½ molecular electron and the ½ gold hole wave functions. This depiction is supported by experimental evidence of UV absorbance spectra of a polyalanine film on gold which shows a new absorption band starting at 500nm and progressing to higher wavelengths[10] and for 1- octadecanethiol monolyers on gold[11]. Since at this point there is not enough experimental evidence to determine the exact nature of the new delocalized electronic band, from here on it will be referred to as a plasmon band caused by loosely interacting excitons.

We will first calculate the theoretical number of free excitons in the system. The parameters in consideration are the surface area of each molecule which is known and the sample surface area $0.15(cm^2)$[1]. Dividing the two a number of $3.9 \cdot 10^{13}$ excitons/sample is obtained. This is the theoretical prediction for the number of excitons in the system. Assuming that the absorption results from electron plasma we can calculate the electron density. The relation between the plasma frequency and its density is given by:

$$n_e = \frac{\omega_p^2 \varepsilon \varepsilon_0 m_e}{e^2} \quad (eq.21)$$

Where $n_e$ is the electron density, $\omega_p$ is the plasma frequency, $\varepsilon$ the dielectric constant of the layer, $\varepsilon_0$ the permittivity of free space, $m_e$ is the electron mass and $e$ is its charge. For absorption at 500nm which corresponds to a frequency of $\omega_p = 6 \times 10^{14}$ sec$^{-1}$ and $\varepsilon=2.5$, one obtains a density of $2.8 \times 10^{20}$ excitons/cm$^3$. Substituting for the length of the molecule and the sample size we obtain a density of $1.3 \times 10^{13}$ excitons/sample, a factor 3 smaller than the theoretical prediction. The small deviation from the theoretical prediction can be attributed to pinholes or small degree of disorder in the monolayer.

The exciton acts as a pair of positive and negative charged particles. The lowest energy is the ground state. In the ground state the spins of the positive and negative charge align antiparallel. In an applied magnetic field they will remain in the ground state with the total magnetic vector of their spins aligned parallel to the field direction. Each exciton than, acts as a magnetic source with a magnetic moment of one Bohr magneton, $\mu_B$. The magnitude of magnetic moment for each exciton is the same as predicted for each plasmon mode in thin metallic film[14]. Multiplying by the number of excitons and converting to emu we obtain a total sample magnetization of $M=1.2 \cdot 10^{-7}$ emu which is 2-3 orders of magnitude smaller than the magnetization observed for polyalanine monolayer on gold[1]. It should be realized at this point that the nature of the exciton is of a boson particle, in similarity to the boson particle nature of plasmons in thin metallic films[14].

We realize that a magnetic moment of $1.2 \cdot 10^{-7}$ emu/sample produced by the excitons spins is not sufficient to account for the observed magnetic moment of the films, three orders magnitude larger. Thus, some sort of enhancement mechanism must take place in order to explain the 2-3 orders of magnitude increase in magnetic moment observed for organic films on gold. It will be shown in the following paragraph that a single mechanism of vacuum polarization by the conduction excitons in the layer can account for the increase in the monolayer magnetization. This is similar to the diamagnetism of thin metallic films which is produced by plasmons. However, so far theory[14] has underestimated the magnitude for this diamagnetism observed experimentally[12] since enhancement by vacuum polarization has not been considered.

The Casimir effect proposed by Hendrik Casimir[19], and later verified experimentally, along with the lamb shift[20] provides physical evidence for the existence of zero point energy (ZPE). The microscopic version of the Casimir effect is the van der Waals forces in closely spaced atoms and molecules[21]. Quantum electrodynamics (QED) calculations indicate that the speed of light slightly increase when light propagates between parallel conducting plates[22]. In these calculations the vacuum between the conducting plates was treated as a dispersion medium with a refractive index n<1. Due to a change in the permittivity/permeability of the vacuum the speed of light between the two plates increases. The effect is due to a local decrease in the ZPE density (U) between the conducting plates. This doesn't necessarily violate Einstein theory of relativity because the change is too small to be measured within the Heisenberg uncertainty principle, but does imply a change in the fabric structure of the vacuum and the local ZPE density. The change in ZPE energy is too small to account for any of the current results. However, recently it was proposed[14-16] that ZPE can change by orders of magnitude due to interaction of metal plasmons with the vacuum. Since the angular momentum of our system increases by a factor q (eq.9) and time increase by $q^3$ (eq.8) (frequency decreases by a factor $q^3$) the ZPE, a sum over all the vacuum modes $U = \sum_{1}^{n} h\nu/2$ (eq.22) is predicted to decrease by a factor of $q^2$ in agreement with equation-10. We also find that due to electrostatic considerations[23] $\mu_0$ increases by a factor $q^2$ and no change occurs in $\varepsilon_0$. Since $C^2=1/(\varepsilon_0\mu_0)$ it follows that a change in ZPE will cause the speed of light to decrease by a factor q in the area of the gold-monolayer interface. Changes in the speed of light are predicted[14] due to coupling of the photons to the surface plasmons.

It is suggested here, that polarization of the vacuum at the gold-monolayer interface is produced by the new delocalized electronic states as observed in UV absorption spectrum. Changes in physical parameters are directly related to the decrease in the ZPE at the molecular-gold interface, where the degree of change depends on the scaling factor q which is proportional to the inverse of the fine structure constant.

The large magnetization observed for the films can be understood at this point. The magnetic moment of the exciton spin ($1.2 \cdot 10^{-7}$ emu) is enhanced in two ways: 1. Increase in the permeability ($\mu_0$) of the vacuum by a factor $q^2$.
2. Spin Bohr magneton $\mu_B = e\hbar/2m_e$ is increased by a factor q due to an increase in ℏ (eq.9).
Two factors contribute to the magnetic moment of the exciton spin. The first is the magnetic moment of the expanded ½ electron ($6 \cdot 10^{-8}$ emu) that is multiplied by a factor $q^3$ due to 1 and 2. The second is an enhancement of the magnetic moment of the ½ gold hole by a factor $q^2$ due to 1. Combining the two factors we obtain a magnetization of $3.8 \cdot 10^{-5}$ emu/sample, in good agreement with the upper limit of experimental observations[1].

A recent theory has suggested[24] that the giant magnetization originates from boson particles created by triplet pairing of electrons confined in the two dimensional lattice of the monolayer. The magnetization is produced by orbital angular momentum of the triplet pair. An important role in determining the properties of the monolayer is given to bosons in both the former and in the present theory. Yet, the current model attributes the magnetization to the magnetic moment produced by the spins of the exciton pair, enhanced by vacuum polarization. In the ground state of the exciton angular momentum does not contribute to the total magnetization due to L=0.

Interestingly, ideas on tapping the ZPE arise by merging theories of ZPE and system self-organization[25] in our case the formation of the self assembled polyalanine film on gold. Thus, formation of an ordered state by the self-assembly process and the interaction of the monolayer with the gold forms a new state of matter that can interact strongly with the ZPE. The organization and the ordered state of the layer play an important role as to stabilize the formation of the new delocalized electronic state.

Possibly also cooperative effects of Van der Walls forces in the molecular layer enhanced by the plasmon field might contribute to the strong interaction with the ZPE. As suggested by the agreement between the experimental and the theoretical values calculated using the scaling procedure, the strength of the interaction is mediated by the fine structure constant through the relation $\alpha$ has with the vacuum. The proposed mechanism for the large increase in the ZPE is polarization of the vacuum by the large, highly polarizable charge density of excitons - $9 \cdot 10^{13}$ excitons/cm$^2$ confined in the small area of the gold-monolayer interface.

The large spin selectivity of the film, the CPD, its temperature dependence, the appearance of resonances and the enhanced magnetization, are all linked to a change in physical parameters of the gold-monolayer interface that depend on the fine structure constant. The value of the fine structure constant represents the relationship between the energy stored in electromagnetic interaction of charged particles and the energy associated with their mass. The formation of a monolayer from high dipole molecules creates an intense electric field that must be decreased in order for the system to reach a stable state. It is proposed that the reduction in the electric field strength (eq.12) by $q^4$ and the energy of the electromagnetic modes (eq.22) by a factor $q^2$ is achieved by polarization of the vacuum, were the magnitude of polarization is proportional to the inverse of the fine structure constant. The decrease in the energy of the electromagnetic modes in the vacuum does not modify the value of $\alpha$ since the energy associated with the mass of the particle $E=mc^2$ decrease by a factor $q^2$ as well (eq.5). Applying the scaling factor (as demonstrated previously for the permittivity and permeability) to the fine structure constant, $\alpha = e^2/4\pi\varepsilon_0 \hbar c = \mu_0 c e^2/2h$, shows that indeed $\alpha$ does not change its value. Although at this stage the role of $\alpha$ is purely suggestive further study will be required to exactly determine the relation between the fine structure constant and the appearance of this unique new type of magnetism.

**Epilogue:** This manuscript has been written around 2006. During that time I have realized two important things:

1. The arbitrary factor 2 in Eq.2 that was used originates from Dirac magnetic monopole. This leaves the factor 68.5 which is $q^2$, q therefore is the radius of curvature and $q^2$ the area of the curvature.
2. The physical meaning of the fine structure comes from the geometry of space-time. The electron samples extended areas of past and future making the events of past and future coherent. This makes the system more coherent and results in extension of the wave nature of the electron and appearance of the fine structure constant. Further coherency in space phase by propagation in curved space that causes a reduction in the speed of light, originates from the fact that in the scaling procedure according to Kepler, space scales by $q^2$ while time by $q^3$ and the ratio outcome is reduction in the speed of light by the factor q. That is, the scaling procedure here results from the fact that in curved space, the dimension of time is larger than of space. In the current example of magnetism induced by vacuum polarization the fine structure constant does not change since both electromagnetic fields energy and energy of the particle are reduced, but the effect is quantized in $\alpha$ units. This is caused by change in the fabric of space induced by the interaction of the hybrid system with the vacuum. A different view is that the life time of the virtual electron positron pair of the vacuum is increased and they are present in existence for longer period, the reason why Plunk constant increase and speed of light decrease. The life time of the positron electron virtual particles which determines the coherent length of the system now increase. Instead of the electron radius in the oxygen atom which is the distance light travels without the electronic system disturbed, the light has longer time to travel until the electron wave function is disturbed (the coherence time) therefore, the wavelength can expand and reach the original distance 0.48Å now multiplied by $q^2$.

    It is predicted here that experiments on Casimir forces in such systems should revile large deviations of the near surface ZPE of self-assembled monolayer compared with bare metal surface.

References:

[1] I. Carmeli, G. Leitus, R. Naaman, S. Reich, Z. Vager, *J. Chem. Phys*. **118**, 10372 (2003).
[2] I. Carmeli, Z. Gefen, Z. Vager and R. Naaman, *Phys. Rev. B.* **68**, 115418 (2003).
[3] I. Carmeli, G. Leitus, R. Naaman, S. Reich, and Z. Vager, *Isr. J. of Chem.* **43**, 399 (2003).
[4] S.G. Ray, S.S. Daube, G. Leitus, Z. Vager and R. Naaman, *Phys. Rev. Lett*. **96**, 036101 (2006).
[5] A. Hernando et. al. *Phys. Rev. B*. **74**, 052403 (2006).
[6] P. Crespo et.al. *Phys. Rev. Lett*. **93**, 087204 (2004).
[7] R. Naaman and Z. Vager, *Phys. Chem. Chem. Phys*. **8**, 2217 (2006).
[8] M. Ventakesan, C.B. Fitzgerald and M. D. Coey, *Nature*, **430**, 630 (2004).
[9] G. Kopnov, Z. Vager and R. Naaman, *Adv. Matt*. **19**, 925 (2007).
[10] I. Carmeli, "The effect of spin on photoelectron transmission through organized organic thin films" Ph.d thesis, Weizmann Institute of science, Rehovot, Israel, page 17-18 (2003).
[11] O. Neuman and R. Naaman, *The Journal of Phys. Chem. B. Lett*. **110**, 5163 (2006).
[12] S. Reich, G. Leitus and Y. Feldman, *Appl. Phys. Lett*. **88**, 222502 (2006).
[13] R. Deblock, R. Bel, B. Reulet, H. Bouchiat and D. Mailly, *Phys. Rev. Lett*. **89**, 206803 (2002).
[14] I. I. Smolyaninov et.al. *Phys. Rev. B*. **71**, 035425 (2005).
[15] I. Yoseph, *Phys. Rev. Lett*. **95**, 080404 (2005).
[16] I. Norio, *J. Phys. Soc. Japan*. **75**, 084004 (2006).
[17] *Berkeley Physics Course*, Vol **4**, pg-33, education development center, Inc. Newton, Massachusetts (1971).
[18] Zeev Vager and Ron Naaman, *Chem. Phys*. **281**, 305 (2002).
[19] H.B.G. Casimir, *Proc. Kon. Ned. Akad, Wetenschap*, **51**, 793 (1948).
[20] W.E. Lamb, *Rep. Prog. Phys*, **14**, 19 (1951).
[21] H.B.G. Casimir and D. Polder, *Phys.Rev*, **73**, 360 (1948).
[22] Scharnhorst, K, *Phys. Lett. B*, **236**, 354 (1990). G. Barton and K. Scharnhorst, *J. Phys.A: Matt. Gen,* **26**, 2037 (1993).
[23] The electrostatic equations are: $\varepsilon_0 = C/m \; (e^2 N^{-1} m^{-2})$ where C is the capacitance, m-length, e- electron's charge and N is the force, applying the scaling factor to the units above we obtain: $e^2 / (N \cdot q^{-4}) \cdot (m \cdot q^2)^2 = e^2 N^{-1} m^{-2}$ thus, there is no total change in the units of $\varepsilon_0$. For the permeability: $\mu_0 = N/A^2 \; (Ns^2 e^{-2})$ where A stands for Ampere and s for seconds. $\mu_0$ increase by a factor of $q^2$: $(N \cdot q^{-4}) \cdot (s \cdot q^3)^2 / e^2 = (N \cdot s^2 \cdot e^{-2}) \cdot q^2$
[24] Zeev Vager and Ron Naaman, *PRL*, **92**, 0872051 (2004).
[25] G. Nicolis, I. Prigogine, SELF ORGANIZATION IN NONEQUILIBRIUM SYSTEMS, Wiley, N.Y. 1977; A. Hasegawa, "Self-Organization Processes in Continuous Media," Adv. Phys. 34(1), 1-42 (1985).